\documentclass[oldversion]{aa}  
\usepackage{graphicx}
\usepackage{txfonts}
\usepackage{natbib}
\usepackage{times}
\bibpunct{(}{)}{;}{a}{}{,}

\def\0{\phantom0}

\begin{document}

  \title{Direct detection of galaxy stellar halos :  \\ 
  \vspace*{1mm} NGC 3957 as a test case\thanks{Based
  on observations obtained  at the ESO  Very Large Telescope (VLT) in
  the Program 077.B0145(A)}}

  \titlerunning{Direct detection of galaxy  stellar halos}

  \author{  P. Jablonka\inst{1,2,3} \and M. Tafelmeyer\inst{1}
  \and F. Courbin\inst{1} \and A.M.N Ferguson \inst{4} }

  \institute{
   Laboratoire d'Astrophysique, Ecole Polytechnique F\'ed\'erale
   de Lausanne (EPFL), Observatoire, 1290 Sauverny, Switzerland
   \and
   Universit\'{e} de Gen\`{e}ve, Observatoire, CH-1290 Sauverny, 
   Switzerland
   \and
   Observatoire de Paris, CNRS-UMR8111, Place Jules Janssen 92190 Meudon, France
   \and 
   Institute for Astronomy, University of Edinburgh, Blackford Hill, Edinburgh, EH9 3HJ, UK
   }

  \date{Received ... ; accepted ...}

\abstract{We present a direct detection of the stellar halo
  of the edge-on S0 galaxy NGC~3957, using ultra-deep VLT/VIMOS $V$
  and $R$ images.  This is achieved with a sky subtraction strategy
  based on infrared techniques.  These observations allow us to reach
  unprecedented high signal-to-noise ratios up to 15 kpc away from the
  galaxy center, rendering photon-noise negligible. The 1$\sigma$
  detection limits are $R$ = 30.6~mag/arcsec$^2$ and $V$ =
  31.4~mag/arcsec$^2$.  We conduct a thorough analysis of the possible
  sources of systematic errors that could affect the data:
  flat-fielding, differences in CCD responses, scaling of the sky
  background, the extended halo itself, and PSF wings.   We conclude
    that the $V-R$ colour of the NGC~3957 halo, calculated between
    5 and 8 kpc above the disc plane where the systematic errors are
   modest, is consistent with an old and preferentially
    metal-poor normal stellar population, like that revealed in
    nearby galaxy halos from studies of their resolved stellar
    content.  We do not find support for the extremely red colours found
    in earlier studies of diffuse halo emission,
    which we suggest might have been due to residual systematic
    errors.  }{ }

\maketitle

\section{Introduction}

Galaxy stellar halos contain fundamental clues about galaxy assembly.
$\Lambda$CDM models predict stellar halos to be ubiquitous and
dominated by old metal-poor populations, characterized by significant
substructures in the form of tidal debris from accreted satellites
\citep{bullock05,abadi06,font06}.  Conversely, in classical
dissipative collapse models, halos are expected to exhibit smooth age
and metallicity gradients \citep[e.g.,][]{1962ApJ...136..748E}.  Due
to the extreme faintness of these envelopes, the value of galactic
stellar halos as key tests of galaxy formation theories has however
not yet been fully realized.

Since the middle of the nineties, our view of galaxy haloes has
changed dramatically with the discovery of substructures around the
Milky Way \citep[e.g.,][]{ibata94, 1999AJ....118.1709M, newberg02,
  2002ApJ...574L..39G,rochapinto03, 2006AJ....132..714V,
  2006ApJ...645L..37G, 2007ApJ...658..337B, 2008ApJ...673..864J,
  2008ApJ...680..295B}.  Similar signatures of tidal destruction of
satellites by their massive hosts were found in the halo in M31
\citep[e.g.,][]{ibata01,2002AJ....124.1452F,ibata07,
  richardson08,2009Natur.461...66M, 2010ApJ...708.1168T}. Stellar
  streams were also detected around NGC 5907 and NGC 4013
  \citep{2008ApJ...689..184M,2009ApJ...692..955M}. It is nevertheless
  not yet totally clear how ubiquitous halos are and how often mergers
  intervene in their building-up.

Resolving individual stars is mainly limited to a small sample of
galaxies within or close to the Local Group, with a few exceptions.
\citet{2005ApJ...633..821M,2005ApJ...633..828M} resolved the
stars near the red giant branch tip in the outskirts of eight nearby
galaxies with the HST/WFPC2, although it remains unclear for some of
their fields whether they sampled the galaxy outer discs or  halos. 
  Similarly, \citet{2007MNRAS.381..873M,
    2009MNRAS.395..126I,2009MNRAS.396.1231R} analysed the
  extra-planar stellar populations in NGC 891 using three HST/ACS
  pointings.  Recently \citet{2009AJ....138.1469B} conducted the
first wide-field ground-based survey of the red giant branch
population in the outskirts of M81 using Suprime-Cam on the 8-m Subaru
telescope. They detected a faint, extended structural component beyond
the galaxy's bright optical disc sharing some similarities with the
Milky Way's halo, but also exhibiting some important differences.

Searches for extended low surface brightness diffuse halo emission
were carried out around several more distant external galaxies.
The first detection was announced by \citet{sackett94} around the
edge-on Sc galaxy NGC~5907 and later confirmed by \citet{lequeux96}
and \citet{lequeux98}.  These two subsequent works additionally
reported an unexpected reddening of the halo colours, increasing with
the distance to the galaxy center. These results were contested by
\citet{zheng99}, who noticed that artifacts in the star-masking
procedure could be responsible for the apparent reddening.
\citet{zibetti04} used a stacking technique to combine the images of
1047 edge-on galaxies selected from the Sloan Digital Sky Survey.
They detected the presence of a mean luminous halo, whose shape was
clearly rounder than the disc.  Their data suggested a correlation
between the halo and galaxy luminosities.  Surprisingly, their halo
colours were redder than the reddest known elliptical galaxies.
\citet{zibetti04b} reached similar photometric depths in a study of a
single edge-on galaxy in the Hubble Ultra Deep Field, where they also
found anomalously red colours, not reproduced by any conventional
stellar populations model and getting redder with larger radii. 
  \citet{dejong08} however  discussed how scattered galaxy light
  from extended point spread function tails could have been
  underestimated in these studies and thus lead to spurious
  detections.

The controversy over the existence and nature of stellar halos around
external galaxies reflects  that these types of observations
are extremely challenging from a technical standpoint.  Errors in
flat-fielding and sky subtraction can easily mask or contort real
signal.  In order to overcome these limitations and to understand
galactic halo properties, we designed a new strategy 
  designed for galactic halo surface brightness profiles. It is based
on two principles: {\it i)} we target a galaxy, which has an angular
size that is much larger than the PSF extent and {\it ii)} we carry
out very accurate sky subtraction, with a chopping technique similar
to that used in near-IR imaging.

This paper is organized as follows: Section~\ref{sec:strategy}
presents our observational technique.  Section~\ref{sec:reduction}
describes the data reduction procedure leading to ultra-deep
sky-subtracted $V$ and $R$ images of NGC~3957.
Section~\ref{sec:results} presents the radial surface brightness
profiles derived for the halo of NGC~3957.  Section~\ref{sec:errors}
discusses the most important possible sources of errors, focusing on
the systematics.  Section~\ref{sec:discuss} interprets the $V-R$
colour profile. Our conclusions are summarized in
Sect.~\ref{sec:conclusion}.

\begin{table}[t!]
\caption[]{Journal of the observations. Half of the exposures of each night were taken with the galaxy located in CCD1 and the others with the galaxy in CCD2
(see Fig. \ref{fig:vimos}).}
\label{tab:obs}
\begin{flushleft}
\begin{tabular}{ccc}

\hline\hline
Date & Exposure & Filter\\
(jjjj/mm/dd) & &\\
\hline 
2006/04/24 & 24 x 190s & R  \\
2006/04/25 & 24 x 190s & `` \\
2006/04/27 & 12 x 190s & `` \\
2006/04/28 & 24 x 190s & `` \\
2006/04/29 & 24 x 190s & `` \\
2006/05/25 & 24 x 190s & `` \\
\hline
2006/05/27 & 12 x 450s & V  \\
2006/05/28 & 12 x 450s & `` \\
2006/05/29 & 12 x 450s & `` \\
2006/05/31 & 14 x 450s & `` \\
\end{tabular}			
\end{flushleft}
\end{table}

\begin{figure*}[t!]
\begin{center}
\includegraphics[width=7.cm]{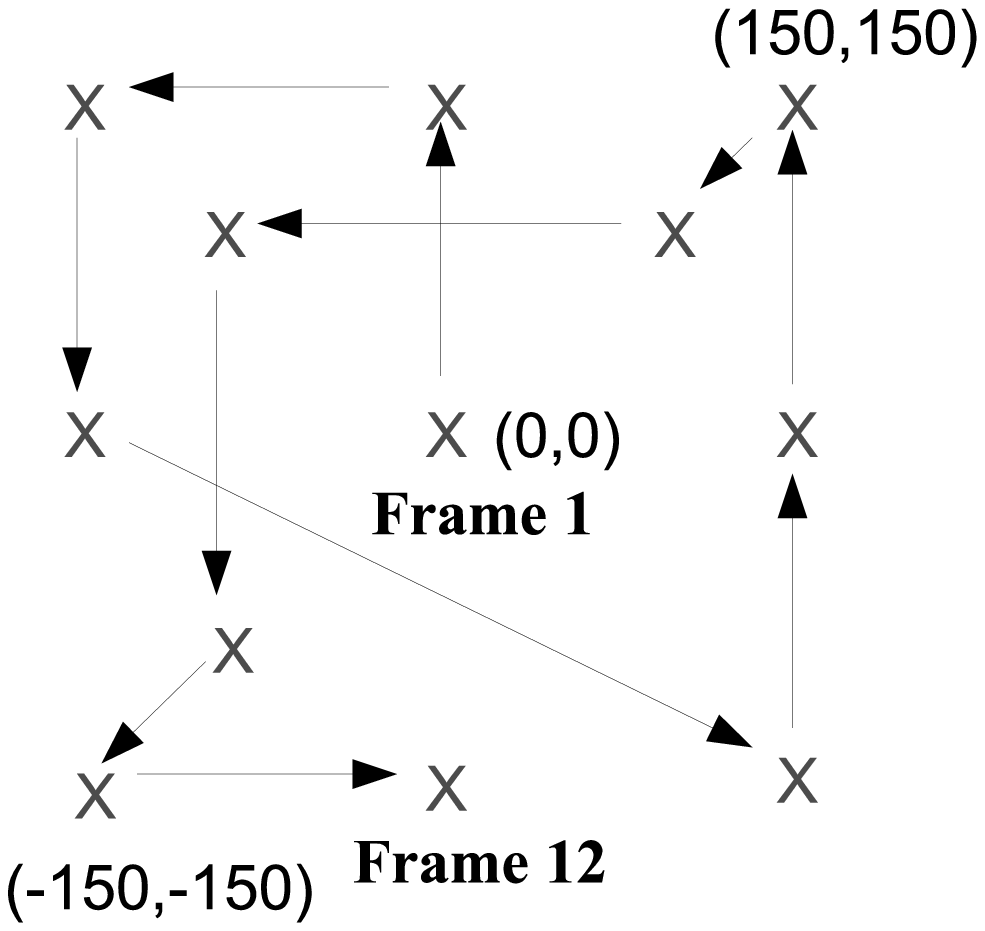}
\hskip 30pt
\includegraphics[width=7.cm]{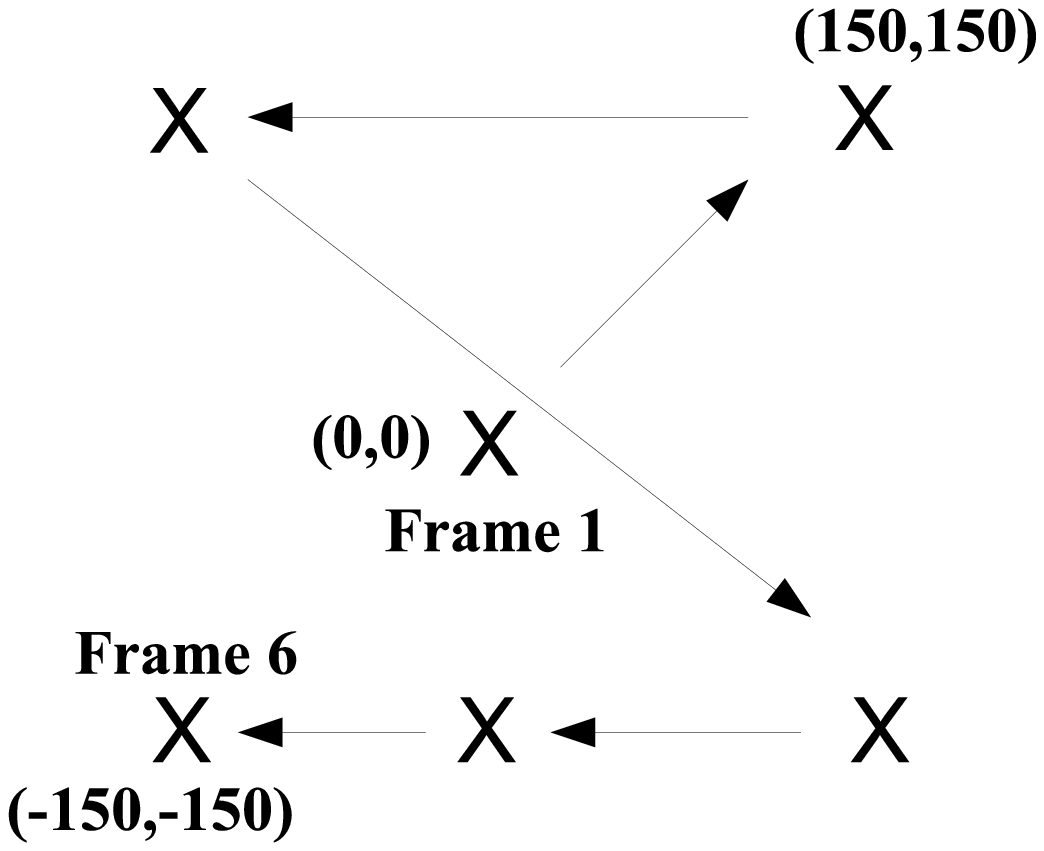}
\caption{{\it Left:} dithering pattern for a set of 12 consecutive $R$-band 
frames. The position (0,0) indicates the initial position of NGC 3957
in one CCD. {\it Right:} dithering pattern for a set of six
consecutive $V$-band exposures. Units are given in pixels. One pixel
is 0.205\arcsec.}
\label{fig:dithering}
\end{center}
\end{figure*}

\section{Observational strategy}
\label{sec:strategy}

\begin{figure*}[t!]
\begin{center}
\includegraphics[width=12cm]{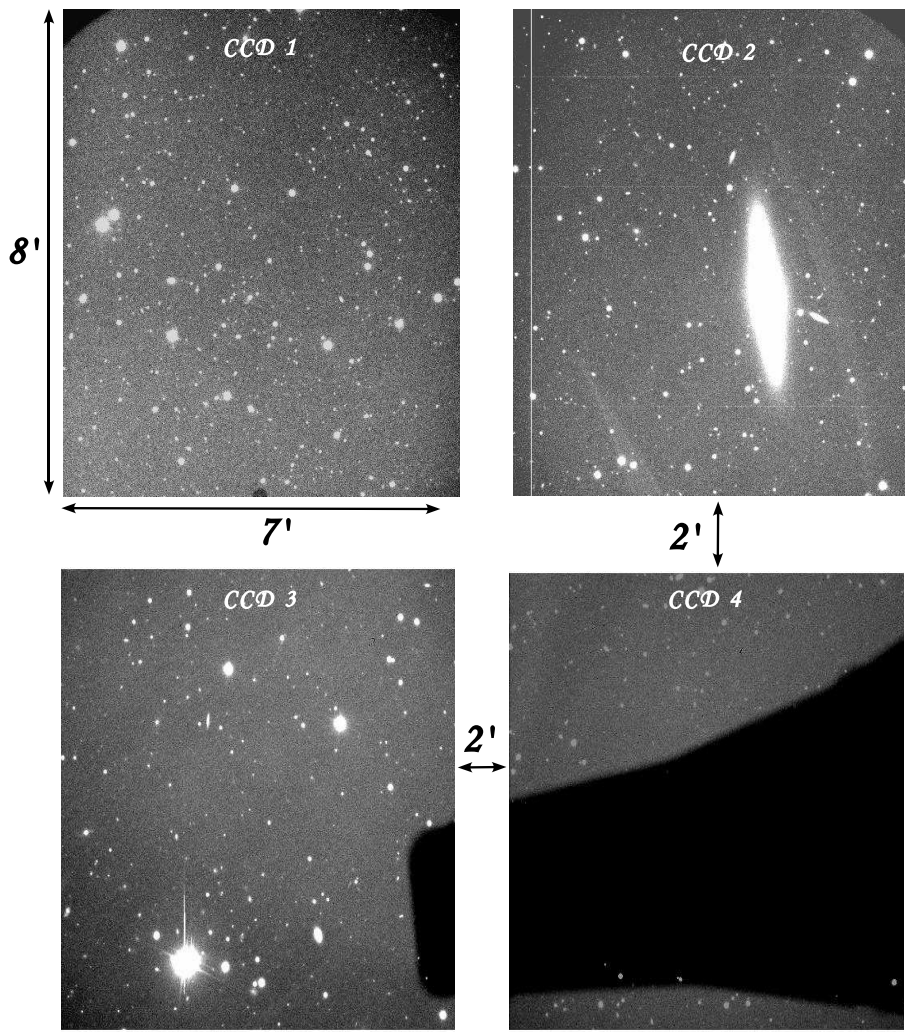}
\caption{Geometry of the VIMOS field of view, composed of four CCDs, prior to
sky subtraction.  The  galaxy was alternately put in CCD1 and in CCD2.  The black feature
  in CCD3 and CCD4 is due to the guide star camera.}
\label{fig:vimos}
\end{center}
\end{figure*}

We selected the early-type (S0) galaxy NGC~3957 (RA(J2000):11 54 01.5
; DEC (J2000) : $-$19 34 08 ; v$_{hel}$=1637 km/s).  NGC~3957 is seen
nearly edge-on, its exact inclination angle (88 $\pm$ 1 degrees) was
measured by \citet{2004A&A...422..465P}. Its apparent size
(3.1'~$\times$~0.7') is large enough to allow us to integrate flux
over large regions when determining the surface brightness profiles. At
the same time, it is small enough to fall entirely in a single VIMOS
CCD chip.  Taking $H_0= 73$ km/s/Mpc, at the distance of NGC~3957
(22.42 Mpc), 1 arcmin=6.52 kpc. Another important point is that the
surroundings of NGC~3957, which were used to determine the background
level and shape, are devoid of any bright objects.  

The choice of VIMOS was motivated by the opportunity to have a wide
field of view distributed over four different CCDs, of 2048 x 2440 pixels
each.  With a resolution of 0.205'' per pixel, each CCD covers 7' x
8'.  They are separated by a gap of 2' (see Fig.~\ref{fig:vimos}).
Unfortunately, the choice of a guide star satisfying VIMOS
specifications had the consequence that the guide probe fell on two of
the four CCDs. In all data, it covers more than 50\% of either CCD3 or
CCD4 (see CCD4 in Fig.~\ref{fig:vimos}), leaving those two chips
unusable for our analysis.

We gathered $R$ and $V$-band images, with total exposures of 22500s
and 25080s (i.e., $\sim$ 6 and 7 hours), respectively.  Exposures were
split into sequences of 190~s ($R$-band) and 450~s ($V$-band).  The
journal of the observations is presented in Table~\ref{tab:obs}.

The design of VIMOS allowed us to place NGC~3957 in one CCD while
integrating the sky on the other one.  This strategy offered two major
advantages: ({\it i}) On one hand, the background level was estimated
in a region located far away from the galaxy, thus avoiding possible
contamination by faint, extended galaxy components.  ({\it ii}) On the
other hand, a 2-D image of the sky was obtained under the exact same
observing conditions as those of the galaxy.

Exposures were dithered following the patterns shown in
Fig.~\ref{fig:dithering}.  This allowed us not only to build master skies
from empty field frames, nicely removing any bright object, but also
to smear out any inhomogeneities in the background, increasing the
flatness of the combined master skies.

As already mentioned, we used two
different pointings.  We  will use the following
terminology throughout:  \textbf{Pointing 1}  corresponds to  a set  of exposures
placing the galaxy   in    CCD1   and   dithered   following    the   patterns   of
Fig.\ref{fig:dithering}.  \textbf{Pointing  2} has the  same dithering
pattern as Pointing 1, but the galaxy is now placed in CCD2. The exposures
on the empty field in CCD1 (when the galaxy is in CCD2) and CCD2 (when
the galaxy is in CCD1) are used to determine the background level 
and shape.

\begin{figure*}[t!]
\begin{center}
\includegraphics[width=12.0cm]{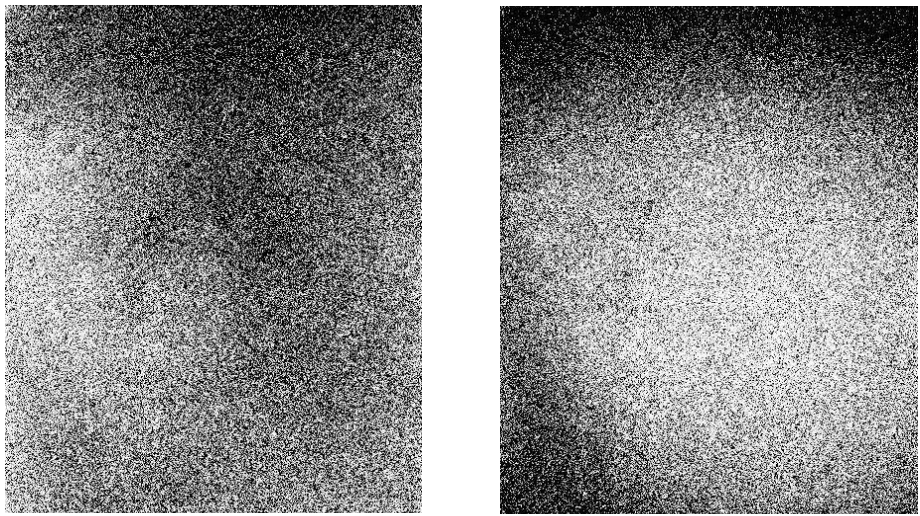}
\caption{Example of the master sky in $R$ (left panel) and $V$ (right panel)
The cuts are $\pm 3 \sigma$, the large scale variations are $\lesssim$1\%.}
\label{fig:msky}
\end{center}
\end{figure*}

\section{Data reduction and sky subtraction}
\label{sec:reduction}

We  only  describe   at length  the  steps  differing from   the classical
methodology for  the reduction of images.  These steps are critical to
reach a    surface   brightness level  down   to    $\sim$  30
mag/arcsec$^2$.

Bias and flat-field corrections were done in a classical manner.
Five featureless biases were taken each night. Their mean ADU/pixel
was subtracted from all frames of the same night.  When possible, a
series of five twilight flat fields were taken.  For each of these
nights, we composed a master flat, corresponding to the average of the
five flats, with a 2-$\sigma$ clipping boundary, $\sigma$ being
calculated with the gain and readout noise of the CCDs. Since
flat-fields were not taken each night, the same master flat was used
for several subsequent nights. We will discuss the possible
uncertainties arising from the flat-fielding in
Sect.~\ref{sec:gradients}.

The most crucial step of the data reduction is the sky subtraction.
One master sky per night and per CCD was created as the average (plus
2-$\sigma$ clipping boundary) of all dithered frames which did not
contain the galaxy. For the $R$-band, 12 sky images per
night and CCD were available.  Sigma clipping of the dithered images
proved sufficient to remove any bright star in the field.  
For the $V$-band, only six frames were available.  Consequently, we had
to mask the bright stars before dithering  to remove them
completely from the master sky frames.  Figure~\ref{fig:msky} provides
an example of a master sky for the two photometric bands.  The cuts
are chosen to be $\pm$2\% of the mean background level, which is three
times the standard deviation.  The large scale structures represent
$\lesssim$1\% of the mean background level.  They are of the order of
the uncertainties in flat-fielding, which are discussed in
Sect.~\ref{sec:gradients}.

\begin{figure}[t!]
\begin{center}
\includegraphics[width=8.5cm]{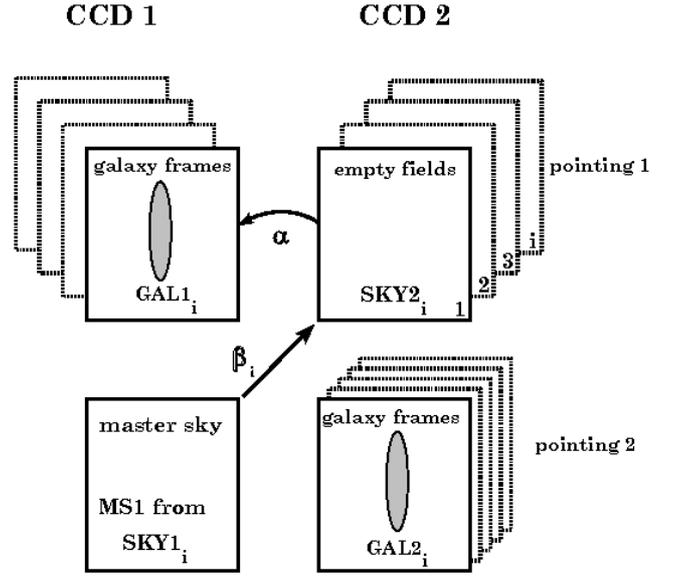}
\caption{Explanation  of  the  scaling  scheme,  prior  to  the  image
 coaddition.   The   master  sky  of   CCD1  (MS1)  is  taken   as  a
 reference. The $i$ index refers to the individual exposures within a
 sequence. The  factors $\beta_{\rm  i}$ scale MS1  to match  the sky
 level of each empty fields of CCD2 (SKY2$_{\rm i}$), observed on the
 same  night.  The  factor $\alpha$  corrects for  the  difference in
 sensitivity  between  the  two  CCDs.   It is  determined  from  the
 intensity ratios of the flatfields of CCD1 and CCD2.}
\label{fig:scaling}
\end{center}
\end{figure}

For the sake of clarity, we will now use the following nomenclature
(see Fig.~\ref{fig:scaling}):

\begin{itemize}
\item \textbf{MS1}: Master sky of CCD1, taken from 
the empty field frames of Pointing 2, 
\item \textbf{SKY1$_{\rm i}$}: individual CCD1  image   of Pointing  2, which does
not contain the galaxy and from  which MS1 was created,
\item \textbf{GAL1$_{\rm i}$}: individual CCD1 image  
of Pointing 1, which contains the galaxy,
\item \textbf{MS2}: Master sky of CCD2 with Pointing 1.  Analogous to MS1.
\item \textbf{SKY2$_{\rm i}$}: individual CCD2 image of the empty field   with
Pointing 1. Analogous to SKY1$_{\rm i}$,
\item \textbf{GAL2$_{\rm i}$}: CCD2 individual galaxy  image  of CCD2 with 
Pointing 2. Analogous to GAL1$_{\rm i}$,
\end{itemize}

The index i goes from 1 to 12 for the $R$-band or from 1 to 6 for the $V$-band.

We could not directly use the sky frames obtained at the same time as
the galaxy.  Indeed, galaxy and background would then have been
observed with different CCDs.  Small differences in the flat-fielding
between CCD1 and CCD2 could introduce artificial intensity gradients
in the final images.  We needed to subtract skies of Pointing 2 from
the galaxy frames of Pointing 1 (and vice versa) obtained during the
same night. In practice, for each night we subtracted
MS1 from GAL1$_{\rm i}$ and MS2 from GAL2$_{\rm i}$, i.e.  the master
skies were observed at a slightly different time than the galaxy, but
with the same CCD.  Although full dithering patterns for both
pointings were completed in about one hour, the sky level was not
necessarily stable in this time interval. Actually, we
measured a variation in the mean background level of 10-15\% from the
first to the last exposure for a given pointing sequence.  Two
corrections were therefore necessary:

(1) For each GAL1$_{\rm i}$ (GAL2$_{\rm i}$), MS1(MS2) had to be
scaled to the level of the sky at the time at which the galaxy was
observed. We had direct access to this information via the
contemporaneous exposures SKY2$_{\rm i}$ (SKY1$_{\rm i}$).  To
determine the scaling factor of each GAL1$_{\rm i}$, MS1 was
multiplied by the ratio

\begin{equation}
\beta_{\rm i1} = \frac{<SKY2_{\rm i}>}{<MS1>},
\label{eq:beta}
\end{equation}

where $<$MS1$>$ is the mean intensity, in ADU/pixel, of a large
central region ($\approx$ 2M pixels, i.e., 22 kpc-radius region
centered at $\sim$50kpc from the galaxy center) of the master sky
frame and $<$SKY2$_{\rm i}>$ is the mean intensity in the same region
of the sky field frame, observed simultaneously with GAL1$_{\rm i}$.  Our
surface-brightness limit is $\sim$ 31.4 mag/arcsec$^2$ in $V$ and
$\sim$30.6 mag/arcsec$^2$ in $R$.  These limiting magnitudes
correspond to the integrated sky noise in  1 arcsec$^2$, or
equivalently, to the fluxes leading to a signal-to-noise ratio of 1
(28.2 mag/arcsec$^2$ in $R$ and 29.0 arcsec$^2$ in $V$ at
S/N=3). \citet{irwin05} report a $V$-band halo surface brightness in
M31 of 31-32 mag/arcsec$^2$ between $\sim$ 40 and 55 kpc along the
minor axis. Since this area of M31 is contaminated by tidal debris
streams, the actual surface brightness of the underlying halo is
somewhat lower (e.g. \citet{ibata07}).  Assuming that the halo of M31
is identical to that of NGC~3957, it corresponds to the level of
noise in our sky frames.  (2)Since $<SKY2_{\rm i}>$ and $<MS1>$ are
measured on two distinct CCDs, we had to correct $\beta$ for their
different responses. It was done by calculating the ratios of the
intensities of the bias-subtracted, non-normalised flatfields of CCD1
and CCD2. Namely, each $\beta_{\rm i}$ had to be multiplied by a
sensitivity correction

\begin{equation}
\alpha = \frac{<FF1>}{<FF2>},
\label{eq:alpha}
\end{equation}

for the scaling of the  MS1 frames and by $\alpha^{-1}$ for the  scaling of MS2.

Errors in $\alpha$ have opposite effects in CCD1 and CCD2
respectively, leading to a systematic additive offset in the results
when considering each of the two CCDs separately. This can be used to
fine-tune $\alpha$ removing any systematic difference between CCD1 and
CCD2.

The complete sky subtraction is finally described by

\begin{equation}
FIN1_{\rm i} = GAL1_{\rm i} - MS1\cdot \alpha \cdot \beta_{\rm i1}
\end{equation}

and

\begin{equation}
FIN2_{\rm i} = GAL2_{\rm i} - MS2\cdot \frac{\beta_{\rm i2}}{\alpha},
\end{equation}

where FIN1$_{\rm i}$ and FIN2$_{\rm i}$ are the final sky subtracted frames
of CCDs 1 and 2. A schematic view of the scaling method is 
shown in Fig.~\ref{fig:scaling}.

The 132 (50) $R$-band ($V$-band) background-subtracted images were
finally aligned and combined, and the final images are calibrated in
the Johnson-Cousins photometric system. One CCD frame per night was
calibrated in $V$ and $R$ with the standard stars observed during
the same period.  The final $V$ and $R$ combined images were then
scaled to the level of their respective reference $V$ and $R$ frames.
Hence, the final frames were multiplied by the factor
flux(single)/flux(combined), where flux(single) and flux(combined)
were measured from several bright stars in the reference and combined
frames, respectively.  These photometrically calibrated images were
then corrected for extinction using $A_{R}=0.124$ and $A_{V}=0.153$
following \citet{schlegel98}.  The extinction varies by 0.006 mag
  rms within a radius of 6 arcmin around NGC 3957.  Considering
random errors alone, the 1-$\sigma$ final accuracy of our calibration
is 0.022~mag, in $R$, and 0.036~mag, in $V$.

\section{Surface brightness profiles}
\label{sec:results}

\begin{figure}[t!]
\begin{center}
\includegraphics[width=8.5cm]{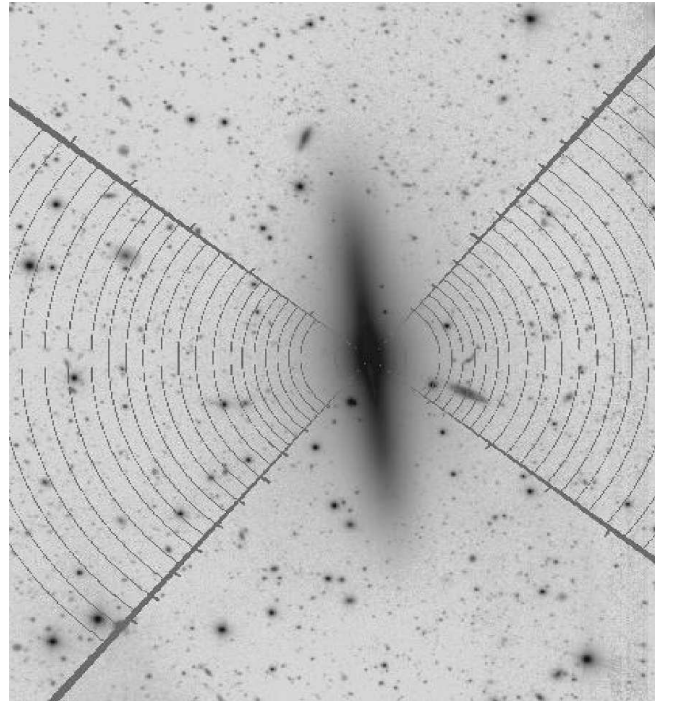}
\caption{Cone shaped regions used to  integrate the light on the final
 coadded $V$ and $R$ images. North is to the top, east to the right.}
\label{fig:cones}
\end{center}
\end{figure}

\begin{figure}[t!]
\begin{center}
\includegraphics[width=8.5cm]{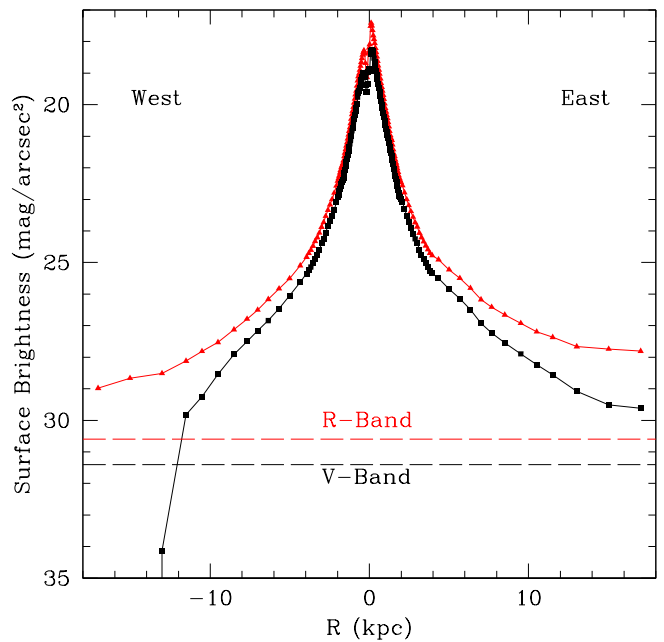}
\caption{$V$- (black) and $R$-band (red) surface brightness profiles
  along the minor-axis of NGC~3957. Errors due to photon noise are
  typically $\leq$0.05 mag within 10~kpc and $\leq$0.1~mag for
  r~$\geq$~10~kpc, hence too small to be plotted.  The systematic
  errors are not displayed for clarity and can be seen in Fig.
  \ref{fig:magfit3}. The dashed lines indicate the 1$\sigma$ detection
  limits in the two bands.}
\label{fig:profiles}
\end{center}
\end{figure}

In order to trace the galaxy signal down to very faint levels with a
sufficiently high signal-to-noise ratio, we integrated flux in
$90^{\circ}$ wedges, running along the minor axis of the galaxy
(Fig.~\ref{fig:cones}).  Prior to this binning, the foreground stars
were carefully masked, as the background galaxies.  The radial
width of each bin was chosen as a trade-off between the spatial
resolution of the galaxy profile and the desired signal-to-noise per
bin.  The narrowest bins have an annular width of 2 pixels (0.04~kpc)
and are located in the central parts of the galaxy, while the widest
ones have a radial extent of 100 pixels (1~kpc). The resulting surface
brightness profiles in $V$ and $R$-bands, obtained by taking the
average of the flux in all unmasked pixel in each radial ring, are
shown in Fig.~\ref{fig:profiles}, where the dashed lines indicate the
detection limits per square arcsecond.  The spatial sampling is five
times finer than that of \citet{zibetti04} in the inner galactic regions,
while in the outer parts, the spatial sampling of the two studies is
the same. However, unlike \citet{zibetti04}, who get a signal-to-noise
of 1 for the outermost bins at $\sim$12-14~kpc, we are not limited by
photon noise in these parts (S/N$\sim$20).  Systematic errors (see
Sect. \ref{sec:errors}) start dominating over photon noise at
r=8-10~kpc, where our S/N is still $\approx$ 40 in $V$ and $R$ .

We estimated the photon noise $\sigma_{tot}$
per bin by adding in quadrature the photon noise of all N pixels in a
cone-shape bin

\begin{equation}
\sigma_{tot}^{2} = \sum_{i}^{N} \sigma_{i}^{2},
\end{equation}

where the  sum goes over all  non-masked pixels. The  noise per pixel,
$i$, on the sky-subtracted and coadded frame is

\begin{equation}
\sigma_i = \sqrt{\frac{F_i}{g}+\sigma_{sky}^{2}},
\end{equation}

where $F_i$ is the flux in pixel $i$, $g$ is the electron-to-adu
conversion factor, and $\sigma_{sky}$ is the standard deviation of the
pixels in the sky background. This calculation gives typical error
bars of 0.05~mag in regions located at about 10~kpc from the centre.
The error bars remain smaller than 0.1~mag from 10~kpc to the edges of
the frames at 15~kpc. These small error bars are too small to be seen
on Fig.~\ref{fig:profiles}.  An exception is the west-side of the
V-profile.  This is the faintest of the profiles and reaches the
detection limit already at $\sim$12~kpc.  Therefore, the outermost
data points beyond this distance are not considered further.

\section{Possible systematic errors}
\label{sec:errors}

Whilst our detection of the NGC 3957 stellar halo is not limited by photon
noise, it  may still be  affected by a  number of
systematic errors that we now review  and try to quantify.  This is of 
particular importance when trying to interpret the radial changes in 
the colour profiles.

\subsection{Flat fields}
\label{sec:gradients}

Errors in flat-fielding may cause artificial gradients in the sky
background and, in turn, affect the actual halo colour gradient.
Since our skyflats were not taken each night, a possible source of
error is a the temporal variability of the flat-fields over a period
of several nights.  In order to estimate this variability, we compare
our normalized $R$-band flat-field obtained at the beginning of the
observing run (24 April 2006) with the one corresponding to the end of
the observing run (25 May 2006).  The changes during this one month
period do not exceed 1.5\% across the whole field of view.  Since we
used the same flat-fields for the sky and the galaxy frames (see
Fig.~\ref{fig:scaling}), errors in the flat-fields propagate linearly
into errors in the final profiles.  Consequently, a 1.5\% deviation in
the flat-field results in a 1.5\% deviation in the the surface
brightness.  This translates into a maximum change of shape of 0.01
mag, i.e., completely negligible compared with the photon noise.

\begin{figure}[t!]
\begin{center}
\includegraphics[width=8.5cm]{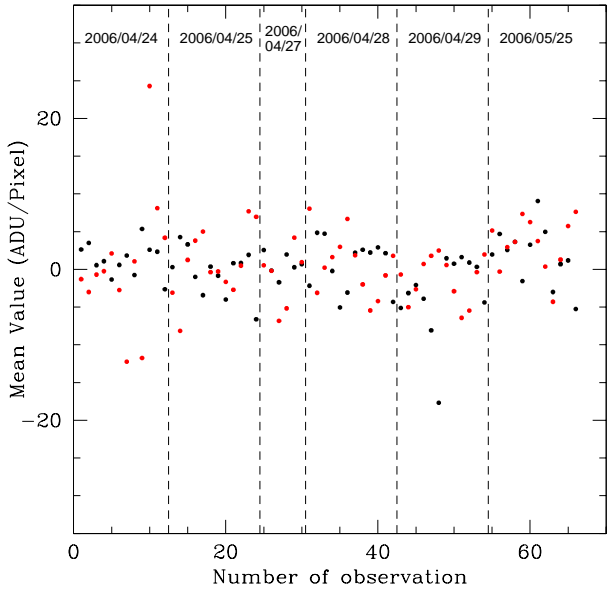}
\caption{Mean flux level as measured in a 100~$\times$~100 pixel empty
  box at the boundary of the individual, sky-subtracted frames.  The
  measurements are for the $R$-band. Black circles are for CCD1, while
  the red circles are for CCD2. There is no systematic difference
  between CCDs, indicating that $\alpha$ is properly determined.}
\label{fig:mean}
\end{center}
\end{figure}

\subsection{Extended PSF tails}

In a recent work, \citet{dejong08} discusses how the extended wings of
the point spread function (PSF) can significantly contaminate the
measurement of galaxy halo light if not properly accounted for.

By targeting a very low redshift galaxy with a size extending over
several arcminutes, we ensured that even the most extended parts of the
PSF did not affect the galaxy's surface brightness profiles at all.
Indeed, the PSF in our coadded frames has a FWHM$\sim$1\arcsec, which
is more than 100 times smaller than the extent of the measured galaxy
halo.  For comparison, \citet{dejong08} deals with PSFs that are about
ten times smaller than the observed galaxy.

In addition, we checked that the wings of the PSF have a limited size.
At 1.8~arcsec (corresponding to 0.2~kpc at the distance of NGC~3957)
the flux in our PSF drops to 0.72\% of its central intensity in $R$,
and to 2.55\% of the central intensity in $V$. The integrated flux of
bright standard stars through apertures of growing radius showed that
there was no measurable flux in the PSF wings already at 15\arcsec\,
away from the core.  Finally, we scaled the PSF to the central
intensity of NGC~3957 and measured the flux in its wings. At 50 pixels,
i.e., 1.1 kpc from the PSF centre, the level of the flux was below
0.01ADU, i.e., not measurable. It is clear that the wings of the PSF
have no effect on our surface brightness profiles.

\begin{figure*}[t!]
\begin{center}
\includegraphics[width=2.5in]{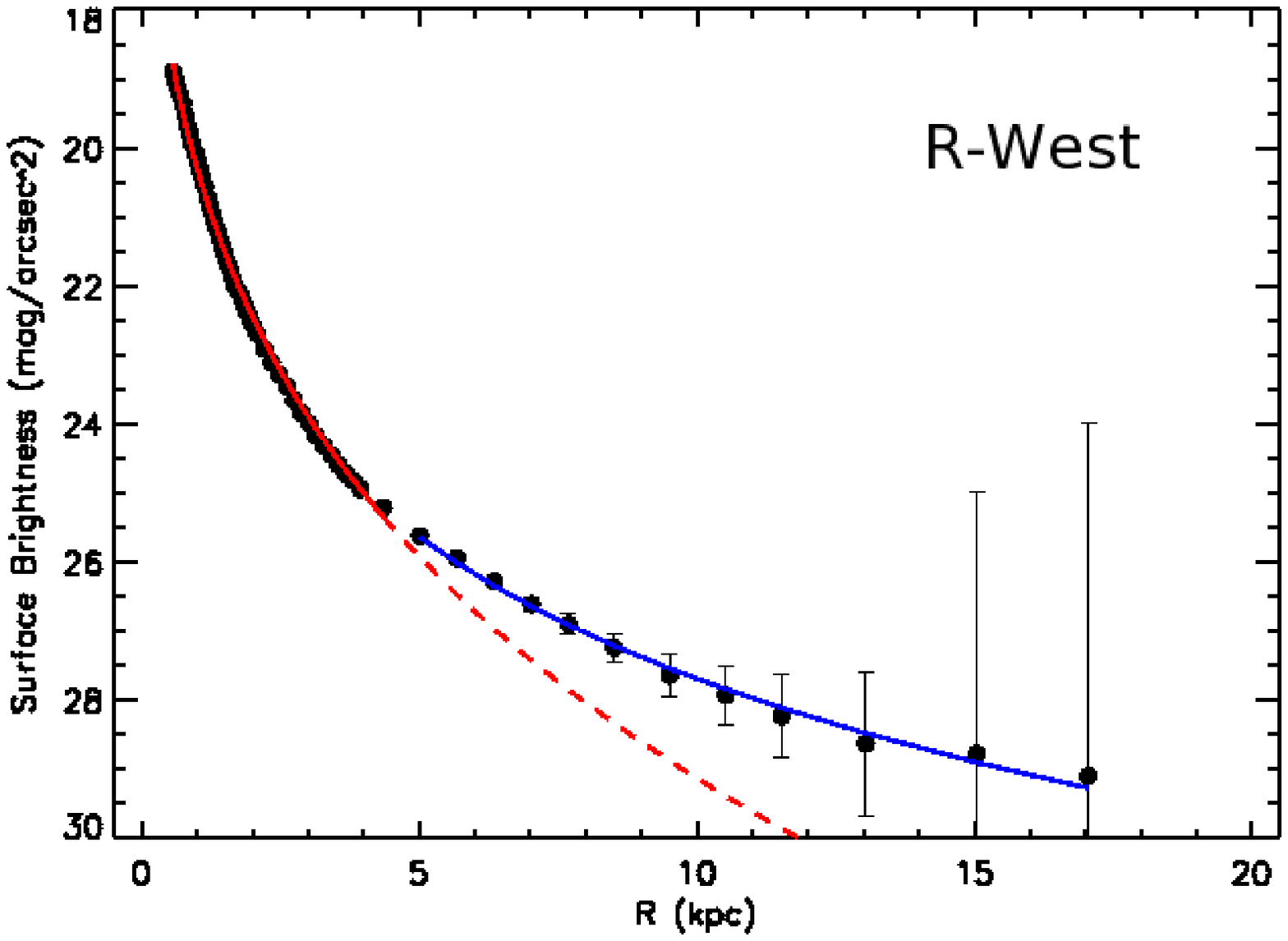} 
\includegraphics[width=2.5in]{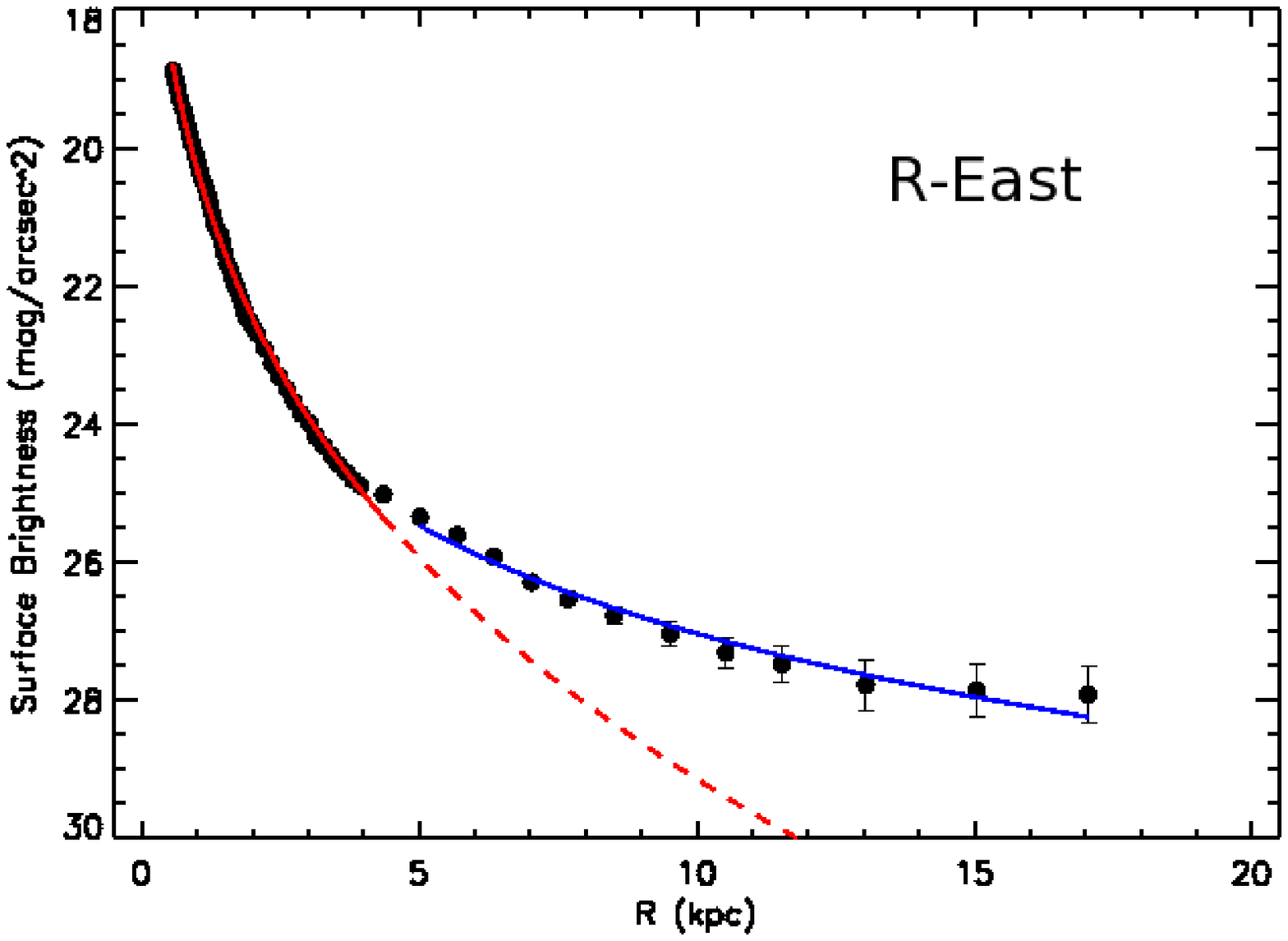}\\
\includegraphics[width=2.5in]{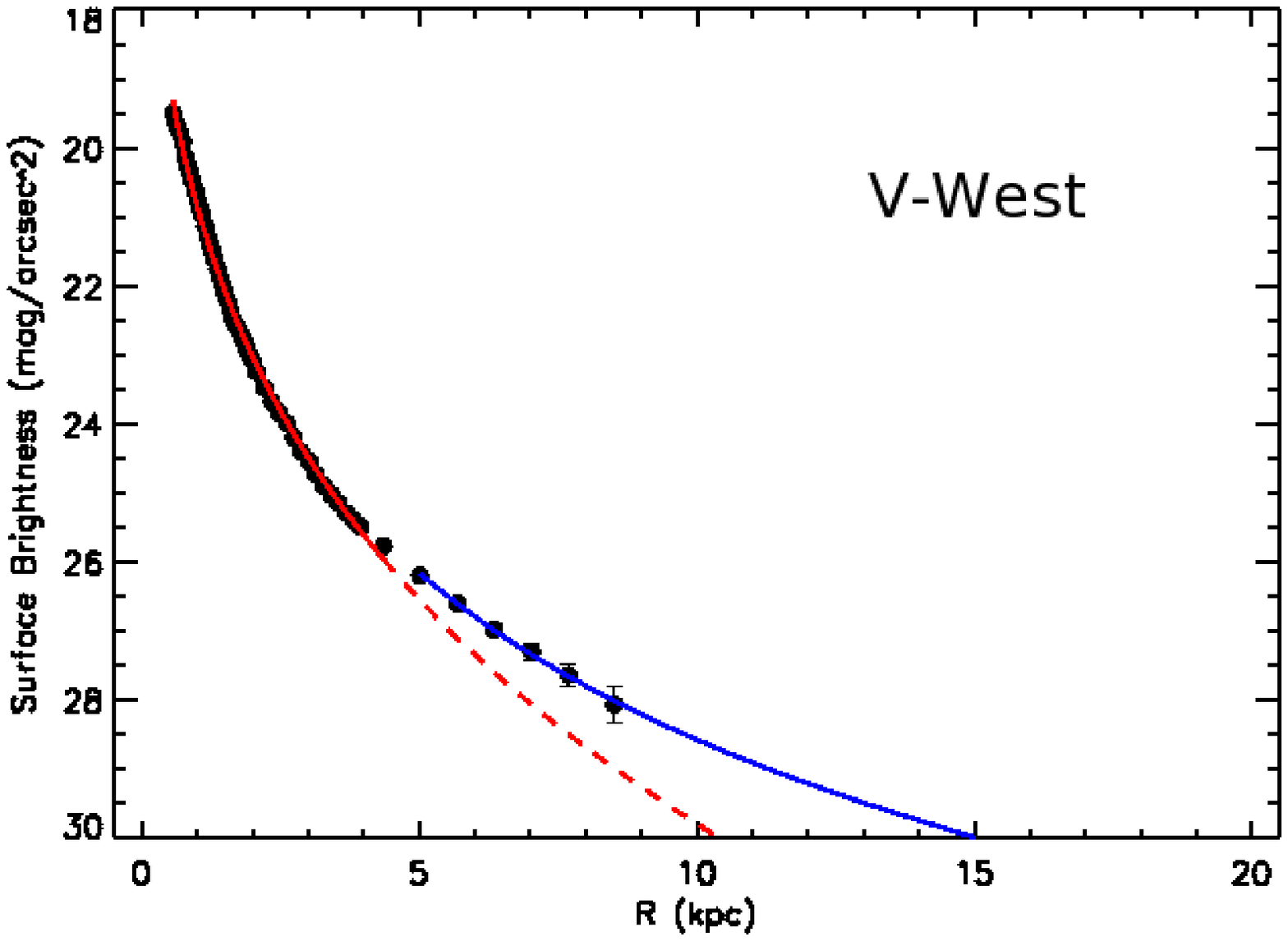} 
\includegraphics[width=2.5in]{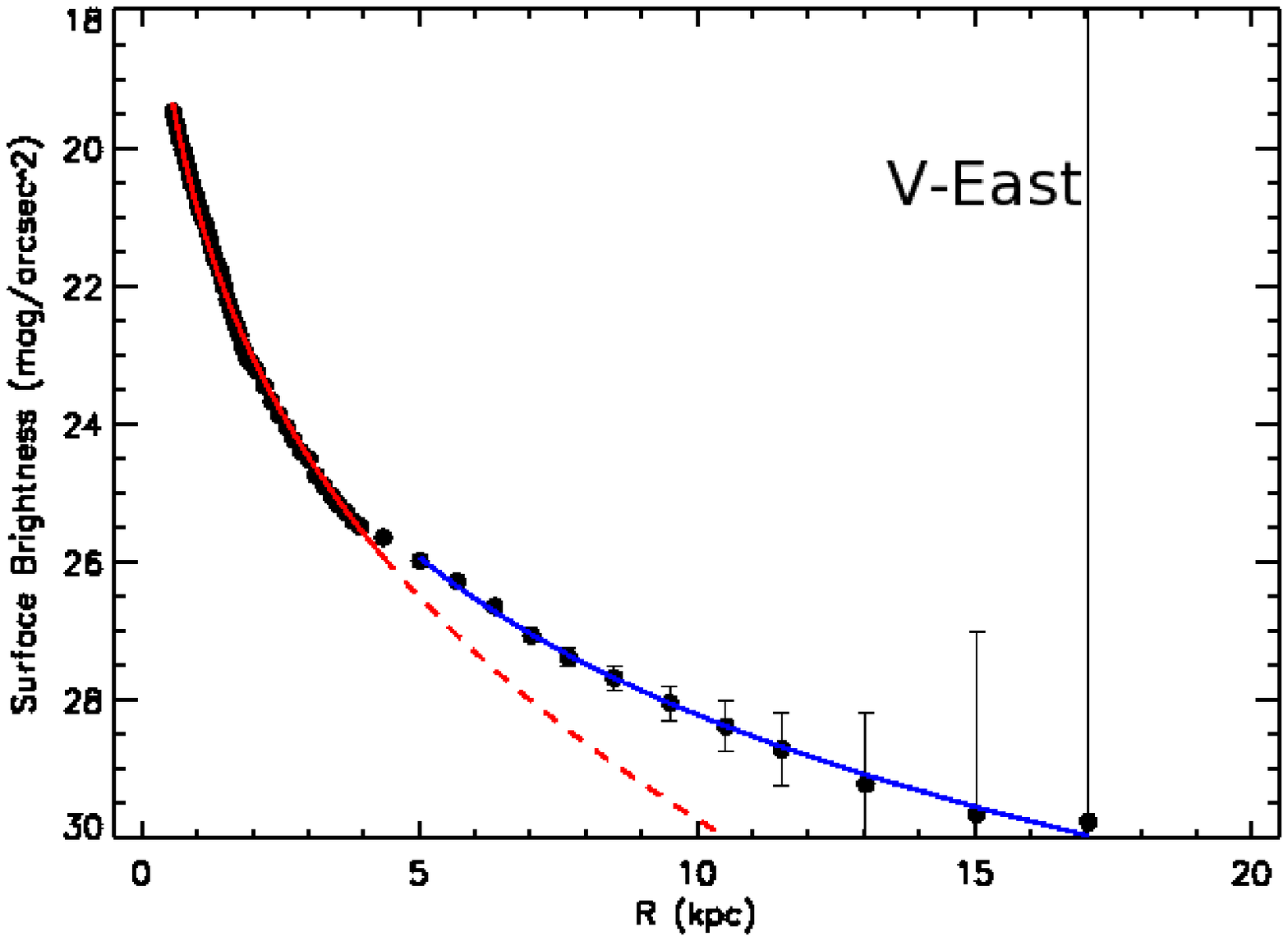}
\end{center}
\caption{The two-component fits to the minor axis profiles in the $V$
  and $R$ -bands.  The observed profiles are shown with solid dots
  with attached error bars (see Sect. \ref{sec:errors}). The last
  point of the V-West profile is not shown, neither is it used in the
  fits, because it is obviously dominated by errors (see Sect.
  \ref{sec:results}). The red lines indicate the de Vaucouleurs law,
  which best fits the bulge of NGC 3957. The red solid line indicates
  the region of the fits within 5 kpc, while the dashed line shows its
  extrapolation to larger radii.  The blue lines show the power-law
  component of the fit at radii larger than 5 kpc.}
\label{fig:magfit3}
\end{figure*}

\subsection{Difference in the CCD responses: $\alpha$-factors}
\label{sec:alphaerrors}

The $\alpha$ coefficient described in Sect.~\ref{sec:reduction}
(Eq.~\ref{eq:alpha}) accounts for the difference in sensitivity
between the two CCDs. The determination of this factor for each
single epoch was carried out in two steps.

1- We initialized a first guess value 
by dividing CCD1 and CCD2 flatfields  by pairs.

2- We applied a small correction to this initial value by noticing that
the scaling of the sky frames subtracted from CCD1 is proportional to
$\alpha$, while the scaling for CCD2 is proportional to $\alpha^{-1}$.
As a consequence, a slight error in $\alpha$ is immediately seen as a
systematic flux offset between the two CCDs, which is minimal when the
value of $\alpha$ is such that the sky subtracts equally well in CCD1
and in CCD2.  In order to evaluate this offset and to correct for it,
we determined in each individual sky-subtracted frame the mean flux
level in a $100 \times 100$ pixels empty region close to the image
boundary.  Our final measurements are shown in Fig.~\ref{fig:mean}
after correction; there is no residual offset between the mean
$\alpha$ factors of CCD1 and CCD2 .

The amplitudes of the corrections applied to $\alpha$ were of the order
1\% at most.  This translates into changes of about 0.01~mag/arcsec$^2$
in the central regions of the galaxy and to 0.1~mag/arcsec$^2$ beyond
13~kpc. The maximum change in slope of the profile, due to an
erroneous estimate of $\alpha$, is therefore at most
0.1~mag/arcsec$^2$ in the faintest and most external regions of the
galaxy field.

\subsection{Sky-level scaling: $\beta$-factors}
\label{sec:betaerrors}

The determination of the mean sky level in each individual image is
affected by shot noise, which in turn leads to an over- or
under-subtraction through erroneous $\beta$ factors
(Eq.~\ref{eq:beta}). While quantifying this error for each image taken
separately is impossible, we used the full data set  to
estimate the amplitude of the fluctuations with time.  This is
equivalent to measuring the scatter of the $\beta$ factors for a
fixed $\alpha$.

The amplitude of the random fluctuations of the scale factor $\beta$
can be estimated from the scatter of the points in
Fig.~\ref{fig:mean}.  The standard deviation of the 132 $R$-band
images is 4.66~ADU, while the uncertainty on the mean $\beta$ is
0.41~ADU. Similarly, the standard deviation of the 50 $V$-band frames
is 3.75~ADU, while the uncertainty of the mean value of $\beta$ is
0.53~ADU. These values provide good estimates of an upper limit on the
errors on $\beta$ in $V$ and $R$.

A systematic error on $\beta$ may have important consequences on the
colour profile of the galaxy, in particular if the amplitude of the
error is different in $V$ and in $R$.  In order to estimate this
effect, we artificially introduced an over- or under-subtraction of the
sky by 2$\sigma$ in both bands. This corresponds to a shift of
2~$\times$~0.41~$\approx$~0.8~ADU in $R$ and
2~$\times$~0.53~$\approx$~ 1.1~ADU in $V$. 

It translates into a
variation of $\sim$0.05~mag at 25.5 mag.arcsec$^2$, $\sim$0.2~mag at 
27 mag.arcsec$^2$ and $\sim$0.5~mag at 28 mag.arcsec$^2$ in the $R$-band.
Similarly, the shifts in the $V$-band 
are $\sim$0.05~mag, $\sim$0.4~mag, and more than $\sim$1.0~mag
at 26, 28, and 30 mag.arcsec$^2$, respectively.

\subsection{Extended halo}
\label{sec:haloerrors}

The last possible source of error is the presence of a very extended
halo signal in the SKY1$_{\rm i}$ and SKY2$_{\rm i}$ frames used to
scale MS2 and MS1 before subtraction (see Fig.~\ref{fig:scaling}).
This rescaling indeed assumes that the CCD frames, which do not contain
NGC~3957, are far enough away from the galaxy to be free from any residual
halo light.

Since the reduction and scaling procedures are the same in the $V$ and
$R$ filters, the systematics causing a positive (negative) shift in
$V$ would also cause a positive (negative) shift in $R$. From their
radial minor axis profiles of M31, \citet{irwin05} derive a power law
surface brightness profile following $I(r)\propto r^{-2.3}$, beyond
20~kpc.  We used this relation to evaluate to what extent SKY1$_{\rm i}$ and
SKY2$_{\rm i}$ might be contaminated by extended halo light from one
CCD on the other. Our sky frames are located at a minimum of twice the
distance from the galaxy centre to the edge of our galaxy profile.
Assuming the \citet{irwin05} M31 halo surface brightness can be
considered representative, the halo light in the CCDs we used to
measure the sky should be at least five times fainter than at the edge of
the CCD that contains the galaxy.

This translates into an error of 1~ADU/pixel, which is a very
conservative value, corresponding to about 40\% of the measured
$R$-band flux and 60\% of the flux in the $V$-band at 13 kpc from the
galaxy centre.  This modifies the $V$ and $R$ East-side magnitudes by
$\sim$0.03~mag at 5~kpc, $\sim$0.2~mag at 10~kpc, and
$\sim$0.3-0.4~mag ($R$) respectively $\sim$0.5~mag ($V$) at 15~kpc.

\subsection{Total error budget}
To summarize, the main possible sources of errors that can affect the
$V$ and $R$ surface brightness profiles of NGC 3957 presented here are
{\it i)} the uncertainties related to the scaling of the sky-level
through the $\beta$-factors and {\it ii)} the possible overestimation
of the actual sky background level, caused by contamination by a
genuine extended stellar halo signal in the regions used to build the
sky frames.  This latter factor can only act as sky over-subtraction
and therefore will contribute only to the upper limit of the error
bars.  Our surface brightness profiles are presented in
Fig.~\ref{fig:magfit3}, along with their total error bars.

\section{Properties of the stellar halo}
\label{sec:discuss}

\subsection{Detection in $V$ and $R$}

As shown in Fig.~\ref{fig:profiles}, we clearly detect light up to
about 15~kpc away from the centre of NGC~3957.  In order to evaluate
the structure of this luminous component and, more importantly, to see
whether one can resolve it into more than a single structure, we
performed 1D-fits to the $V$ and $R$ profiles.  We considered only the
photon noise error bars, since they are the only statistically
meaningful quantities to be taken into account in $\chi^2$ procedures.
We used the {\textit {IDL}} routine {\textit {mpfit}}, which solves the
least-squares problem with the Levenberg-Marquardt technique, to
conduct the fits. The inner 0.5kpc region was not considered due to the
presence of the dust lane. We first started with a de Vaucouleur law
form, chosen to best represent the bulge of this S0 galaxy.
Figure~\ref{fig:magfit3} shows that while a r$^{1/4}$ function with an
effective radius of $0.29$ kpc provides an excellent description of
the inner profile in both in $V$ and $R$ of NGC~3957, it considerably
under-predicts the light beyond 4-5 kpc. Beyond 5 kpc, the excess light
is well-represented by a power-law with index $-2.76 \pm 0.43$.  A
pure exponential fit with a scale length of a few kpc always provides
a worse fits, particularly in the $R$-band, for which it was essentially
discarded. \citet{2009MNRAS.395..126I} applied a de Vaucouleur law
with an effective radius of 1.55 kpc to NGC 891's extra-planar light,
but this did not provide a good description of the light detected
here.
 
{\citet{2004A&A...422..465P} present 3D thick/thin disc
  decompositions for a sample of eight S0 galaxies, including
  NGC~3957. They derive an $R$-band vertical scale height for the
  thick disc component of 2.3 kpc, which, by their definition,
  corresponds to a normal exponential vertical scale height of 1.65
  kpc.  In order to compare this value with our own, we needed to
  extract a vertical profile some distance from the minor axis in
  order to eliminate the influence of the bulge.
  Figure~\ref{fig:disc} shows the $V$ and $R$ West and East profiles
  of NGC~3957 extracted at a position 5 kpc along the major axis.
  Fitting the thick disc between 2 and 4 kpc, we found an exponential
  scale height of 1.27 kpc, which agrees well with that found
  by \citet{2004A&A...422..465P}, considering the different ways in
  which the two studies have extracted profiles and their
  different photometric depths and spatial resolutions.  This fit and
  also that from \citet{2004A&A...422..465P} are overlaid in the
  figure.  Figure~\ref{fig:disc} also underscores the need for an
  additional component beyond a thick disc at vertical distances
  greater than $\sim$4 kpc.

\begin{figure}[t!]
\begin{center}
\includegraphics[width=8.5cm]{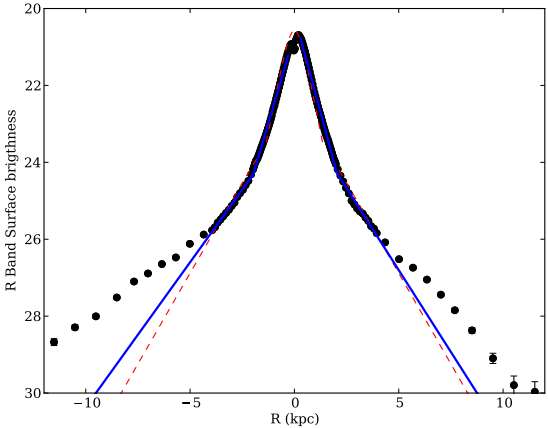}
\caption{The $R$-band vertical profile of NGC 3957 extracted
  at a position 5 kpc along the major axis. The observations are
  displayed with open circles. The \citet{2004A&A...422..465P} fits of
  the thin and thick discs are indicated with a dashed red line, while
  our pure exponential fits are shown with a plain blue line.}
\label{fig:disc}
\end{center}
\end{figure}

\subsection{Colour profile}

Figure~\ref{fig:colour} displays the minor-axis colour profile of
NGC~3957.  The apparent strong reddening at large galactocentric
distances must be analysed in light of the possible systematic errors
presented in Sect. ~\ref{sec:errors}.  As seen in
Fig.~\ref{fig:profiles}, the NGC~3957 surface brightness profile is
traced down to about 1.2 mag above the detection limit both in $V$ and
$R$ bands.

We now analyse the west side of the galaxy (negative values of the
radii) as an illustration of the different elements at play.  The
total error bars in $V$ and $R$ are small and similar from the galaxy
centre, up to about 8.5 kpc ($\Delta \le 0.05$ mag).  At 9.5 kpc, the
error in $V$ is 0.2 mag larger than the one in $R$. This difference
keeps increasing with the radius, reaching 0.95 mag at 10.5 kpc. This
effect is due to the strong decrease in $V$-flux rendering the sky
subtraction very critical, i.e., a given error on the sky subtraction
has much larger impact on the $V-$band fainter profile than on the
$R-$band one. To be quantitative, at 8.5 kpc, the $V$ and $R$ fluxes
are at the level of $\sim$4 ADU/pixel and $\sim$4.5 ADU/pixel,
respectively.  At 9.5 kpc, we measure only $\sim$2 ADU/pixel and
$\sim$3.5 ADU/pixel in $V$ and $R$, respectively.
Figure~\ref{fig:colour} clearly shows that the larger the error bars
due to systematics, the steeper the rise in $V-R$. This suggests that
this reddening may not be a genuine property of the stellar/dust
properties of NGC~3957 halo, but rather  an artifact due to the sharp drop in
the $V$ flux.  The sky subtraction becomes very uncertain and leads to
biased results when the halo surface brightness approaches the
detection limit by less than two magnitudes,
as indicated in Fig.  \ref{fig:profiles}. This is the case at a
distance of 10 kpc from the galaxy centre in $V$, and at 15 kpc in the
$R-$band.

\begin{figure}[t!]
\begin{center}
\includegraphics[width=8.5cm]{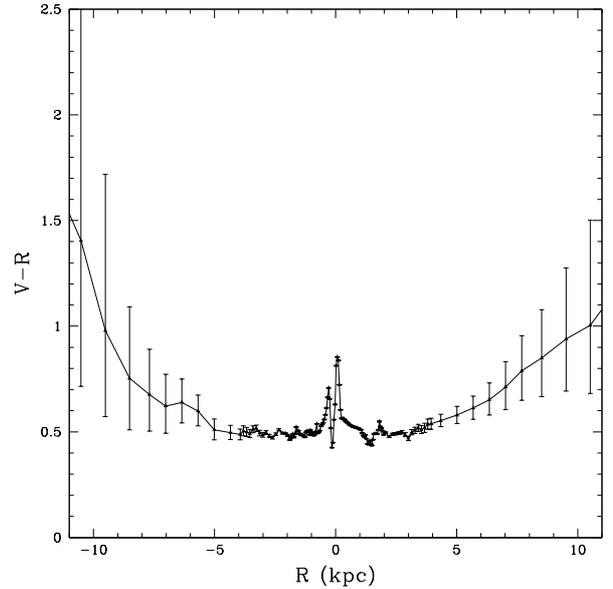}
\caption{The $V-R$ colour profile of NGC~3957. The edges of the vertical
  bars show the maximum possible systematic errors, which dominate
  over the photon noise, as estimated from the uncertainties in the
  sky-level scaling $\beta$ factors (Sect. \ref{sec:betaerrors}) and
  considering possible extended halo (Sect. \ref{sec:haloerrors}) }
\label{fig:colour}
\end{center}
\end{figure}

The west-side mean colour profile has $V-R \sim$0.50 $\pm 0.02$ from
2kpc from the centre of the galaxy, up to 5 kpc and then 
marginally rises to 0.70 $\pm 0.2$, at 8 kpc.  As discussed earlier,
the error bars are not independent from each other and rather reflect
systematic errors.  They all act in the same direction, either
increasing or decreasing the overall $V-R$ profile.  The upper
boundary (i.e., connecting the upper limits of the error bars)
corresponds to an error of 1ADU/pix on the true sky level, or
equivalently to a 3$\sigma$ error (1.2 and 1.6 ADU in $R$ and $V$,
respectively) on the scaling of the sky. Although very modest, these
values have noticeable consequences on the shape of the profiles, due
to the faint luminosity levels considered here. In regions where the
$V-R$ gradient is steeper than 0.1-0.2 mag per kpc, the systematic
errors dominate.  The analysis of the east side of NGC~3957 follows
the same line, albeit the mean colour is redder by
$\sim$0.05 mag with respect to the west-side,  even in the regions where the
systematic errors are negligible.  The small colour
asymmetry between the east and west sides of NGC 3957 is intriguing
and could possibly reflect fluctuations due to substructures within
the galaxy's halo.

We now investigate whether the inner halo colour of NGC 3957 is
indeed compatible with those of normal stellar populations, and
conduct comparisons with a few galaxies for which the halo stellar
population could be resolved and analysed.  We decomposed the halo
population into the same number of single stellar populations (SSPs)
as there are bins in the published metallicity distribution function
(MDF).  These SSPs were generated with metallicities equal to that of
the bin mean using the Y$^2$ isochrones of
\citet{2002ApJS..143..499K}, that extend down to [Fe/H]=$-3.8$, with
[$\alpha$/Fe]=0.3, a Salpeter IMF and a 13 Gyr age. When necessary the
isochrones were interpolated to represent the metallicities of the
MDFs. The resultant SSPs were then summed in luminosity, weighted by
the fraction of light contributed by each MDF bin.

We first derived the expected integrated $V-R$ colour of the Milky Way
halo from the metallicity distribution of
\citet{2009arXiv0909.3019C}. They find that from a vertical distance
of 5 kpc up to 9kpc above the Galactic plane, which corresponds to the
regions we sampled in the present study, the halo stellar
metallicities span [Fe/H]=$-3.0$ to $-1$ with a small fraction of
stars reaching [Fe/H]=$-0.5$. The distributions at 5.5 kpc and 8.5
kpc, peaking at [Fe/H]=$-1.5$ at and [Fe/H]=$-2$, respectively , lead
to $V-R$= 0.49 and 0.47. This narrow range of colours is due to the
progressive insensitivity of the isochrones to changes in metallicity
at low [Fe/H].

As to M31, the complex web of stellar streams prevented us from
deriving a priori a unique colour for its halo. We instead examined the
possible range of values based on earlier works providing complete
metallicity distributions.  The metallicity distribution derived by
\citet{ibata07}  from the diffuse  light of  M31  halo translates into
$V-R \sim 0.51$. The M2 field of \citet{2001AJ....121.2557D}, located
at 20 kpc along the M31 minor axis, peaks at [M/H]$\sim -0.5$ with a
long tail of more metal-poor population down to $-2.5$, leading to an
integrated $V-R \sim 0.54$.  In contrast, the spectroscopic study of
\citet{2008ApJ...689..958K} reveals a strong metallicity gradient 
with a peak at [Fe/H]$\sim -1.$ for galactocentric distances between
20kpc and 40kpc and at [Fe/H]$\sim -2$ beyond 40kpc, with $V-R
\sim$0.47.

Now turning to slightly more distant galaxies
\citet{2007MNRAS.381..873M} analyse the 1.5 first magnitude of the RGB
stars in the halo NGC 891, at $\sim$9.5 kpc from the galactic plane,
centred on $V-I \sim $1.5 and derive a peak metallicity of
[Fe/H]$\sim -1$, i.e, close to an integrated colour of $V-R \sim
0.5$. Some of the galaxies in the sample of
\citet{2005ApJ...633..821M} have metallicity distribution peaks at
[M/H]$\sim -0.6$. Their full metallicity distributions results in an
integrated $V-R \sim 0.54$.

In summary, $V-R$ from $\sim$0.45 to $\sim$0.6 constitutes the range
of expected integrated colours of old and preferentially
metal-poor galactic halos.  The halo of NGC 3957, with $V-R$ from 0.5
to 0.7 mag, between 5 and 8~kpc from the galaxy centre, where
the stellar halo component is clearly dominating and the
systematic errors are still modest, agrees fairly well with these
numbers.  Beyond these vertical distances, the colour reddening
correlates with increasing errors. The order magnitude of the former
being similar or smaller than that of the latter  suggests that the
colour gradient is due to the uncertainties in sky subtraction at very
faint flux levels.

\section{Conclusions}
\label{sec:conclusion}

We  obtained ultra-deep optical VLT/VIMOS images of NGC~3957, a
nearby edge-on S0 galaxy. The total exposure time of six hours in $R$
and 7 hours in $V$ allowed us to reach the limiting magnitudes of $R =
30.6$ mag/arcsec$^2$ and $V = 31.4$ mag/arcsec$^2$ in the Vega
system.

We devised a new observational strategy based on infrared techniques,
which takes advantage of the large field of view of VIMOS.  By
observing NGC~3957 alternatively in two different CCDs, we were able
to create high signal-to-noise and stable sky frames, which were used
to carry out accurate sky subtraction, without any assumption on the
spatial shape and flux level of the sky background.  The observations
allowed us to reach unprecedented high signal-to-noise ratios up 15 kpc
away from the galaxy centre, rendering photon-noise negligible. They
enabled the clear detection of the stellar halo of NGC 3957 at
distances above 5 kpc.

We conducted a thorough analysis of the possible sources of systematic
error that could affect the vertical surface brightness and colour
profiles: flat-fielding, differences in CCD responses, scaling of the
sky background, extended halo, PSF wings. While a r$^{1/4}$ function
with an effective radius of $0.29$ kpc provides an excellent
description of the inner $V$ and $R$-band minor axis profiles of
NGC~3957, it under-predicts the light beyond 4-5 kpc. Above 5 kpc, the
galaxy light profile requires an additional power-law component with
index $-2.76 \pm 0.43$.

 The most secure part of the NGC 3957 halo colour profile falls in
the range $V-R \sim 0.5-0.7$ mag. This colour is compatible with the
properties of nearby galaxy halos as revealed by the investigation of
their resolved stellar populations and with that of the Milky Way.  An
apparent strong reddening is seen in the outer parts (r $\gtrsim$ 8
kpc) of the $V-R$ profile, similarly to earlier works on galaxy halo
surface brightness profiles. However, our analysis indicates that this
reddening may not be a genuine property of the stellar/dust properties
of NGC 3957 halo, and simply reflects the impact of systematic
errors in sky subtraction at extremely faint flux levels.  It is
possible that previously published reports of extremely red colours in
external galaxy halos could have resulted from similar effects.

Future imaging programmes benefiting from larger fields of view ($\sim
0.5-1.0$ degree on one side) will allow the systematics in halo emission
searches to be controlled even further. Existing facilities to carry
out the experiment include the SuprimeCam on the Subaru telescope and
the VST (VLT survey telescope).

\begin{acknowledgements}
  This research used the facilities of the Canadian Astronomy Data
  Centre operated by the National Research Council of Canada with the
  support of the Canadian Space Agency.  This research is partially
  supported by the Swiss National Science Foundation (SNSF).  AMNF is
  supported by a Marie Curie Excellence Grant from the European
  Commission under contract MCEXT-CT-2005-025869.

\end{acknowledgements}

\bibliographystyle{aa}
\bibliography{bibtex}

\end{document}